\documentstyle[12pt,worldsci]{article}
\input{epsf}

\newcommand{\be}{\begin{equation}}
\newcommand{\ee}{\end{equation}}

\begin{document}
\title{ADIABATIC SURFACES FROM THE LATTICE: EXCITED GLUONIC POTENTIALS}
\author{Chris Michael\\
{\em Dept. of Math. Sci., University of Liverpool, Liverpool L69 3BX, UK}}
\maketitle
\setlength{\baselineskip}{2.6ex}

\vspace{0.7cm}
\begin{abstract}

I describe the results from lattice studies of probing the vacuum with 
static colour sources. Topics include
 \begin{itemize}
 \item Twang the flux tube: string models and hybrid mesons.
 \item Charge up the flux tube: adjoint and sextet sources.
 \item Several sources: how does the colour flux get arranged?
 \end{itemize}

\end{abstract}
\vspace{0.7cm}

\section{Introduction}

The non-perturbative sector of QCD is a very interesting area of study. 
Experimental investigation  is difficult since it is not  possible to
make many controlled changes to explore the response. Lattice  QCD
allows a much more comprehensive set of questions to be asked. In 
particular, we can  vary the quark masses. Moreover the simplifying
approximation of  removing quark contributions to the vacuum (the
so-called quenched approximation)  allows phenomenological models to be
tested in this context also. 

The origin of string models in particle
physics is from the  hadronic string that grew out of dual models. It is
worthwhile  to explore the region of applicability of  hadronic
string-like models. The lattice study of the response of the vacuum to 
static colour sources is an ideal testing ground. 
For a pair of sources in the fundamental representation, there has been
much  progress in  extracting the continuum spectrum from the
lattice.  As well as the ground state which has been well known for  a
long time, there are now comprehensive results for  excited gluonic
levels. I  compare these lattice results  with string models and also
discuss the  small separation limit.  

Hybrid mesons are states beyond the quark model in which the  gluonic
degrees of freedom contribute in a non-trivial way. One  of the clearest
ways to study this area theoretically is from  the excited gluonic
potentials. This gives predictions for  hybrid meson spectroscopy with
heavy quarks. A brief aside will cover  the situation with hybrid mesons
containing light quarks: this is of topical relevance since there are
new experimental data on these  hybrid mesons. 

 I cover  a range of other situations which have been studied using 
static sources. These include the bound state spectrum of a single
adjoint source (the gluelump),  the potential between adjoint sources
and between sextet sources and string-breaking.  Some preliminary
results are described of relevance to  baryons: with 3  static quarks. 
In this case one can compare models with minimal flux length to a simple
 sum of two-body interactions.  Leading to a study of  hadron-hadron
interactions, I  present a summary of results for the energy of four
static sources ($Q\bar{Q}Q\bar{Q}$) and some  new results for the 
interaction energy between two B mesons.

 It is worth recalling the weakness as well as the strength of the
lattice approach  to QCD. Any quantity which can be expressed as a
vacuum expectation  value of fields can be extracted straightforwardly
in lattice studies. Thus  masses and matrix elements of operators can be
determined. What is not  so easy is to explore hadronic decays. This is
difficult because the  lattice, using Euclidean time, has no concept of
in and out states. About the  only feasible strategies are to evaluate the
mixing between states of the same  energy - so giving some information
on on-shell hadronic decay matrices, or to make a model 
dependent analysis~\cite{cmdecay} of lattice results.

\section{Gluonic Excitations between Static Quarks in the Fundamental
Representation}

 In a lattice regularisation, static sources in the  fundamental colour
representation  are readily implemented. In the spirit of the heavy
quark effective theory, they  correspond to heavy quarks or anti-quarks.
Thus in the experimental realm  they give insight into $b$ quark
physics. Because of the lattice regularisation, the self energy of
these  static sources is unphysical. Energy differences, however, are
physical  and it is possible to obtain the continuum limit (ie
extrapolate the lattice  spacing to zero).

Before focussing on excitations of the colour flux between static 
quarks, I recall briefly the great insights that have come from the 
ground state potential itself. Some of these topics  are also covered by
other presentations~\cite{other}. The ground-state  potential energy
$V(R)$ between a static quark and antiquark at  separation $R$ has been
measured accurately on the lattice: the continuum limit  can be obtained
with precision in the quenched approximation  when the quark
contributions to the vacuum are neglected. These results  enable
discussion of confinement at large $R$,  of  heavy quark bound states
and provide one way to  determine the running coupling
non-perturbatively. The distribution of colour electric  and magnetic
fields in the flux tube have also been determined from a
lattice~\cite{VRdist}. This again has impact on models  of confinement.
Another insight obtained from the ground state potential is its spin 
dependence - in particular for the long range confining component  which
proves to be `scalar' exchange~\cite{cmspin,VRspin}.

We now consider gluonic excitations  of the potential between  static
quarks: they will play a r\^{o}le in hybrid mesons, for example. An
analysis of the colour  representation of the quark and antiquark is not
useful when they are at different space positions since  their  combined
colour is not gauge invariant. A better criterion is  to focus on the
spatial symmetry of the gluonic flux. The ground state distribution is
symmetric - it is rotationally  symmetric about the separation axis, and
is symmetric about end  to end inversion.  As well as this symmetric
ground state of the colour flux between two  static quarks, there will
be excited states with different symmetries.  

These can be classified, in the continuum, by the group $D_{\infty h}$. 
This classification is well known in molecular applications.  The 
representations are labelled for $J_z=1$, 2, 3  as $\Sigma$, $\Pi$,
$\Delta$ respectively where $z$ is the axis of separation of  the
fundamental sources which are $R$ apart. The $J_z \ne 0$ representations
are two-dimensional. The other labels of the representations  are $g,\
u$ for $CP=\pm 1$ and, for the one-dimensional $\Sigma$ states only, an 
additional $\pm$  label indicating whether  the sate is even/odd under 
reflection in the plane containing the $z$-axis.

 Note that this discussion of excited gluonic configurations between
heavy quarks is very similar  to that of electronic wavefunctions in a
di-atomic molecule. Indeed the  oxygen molecule has contributions from
non-trivial representations  of $D_{\infty h}$.

\begin{figure}[ht!] 
\vspace{11cm} 
\includegraphics{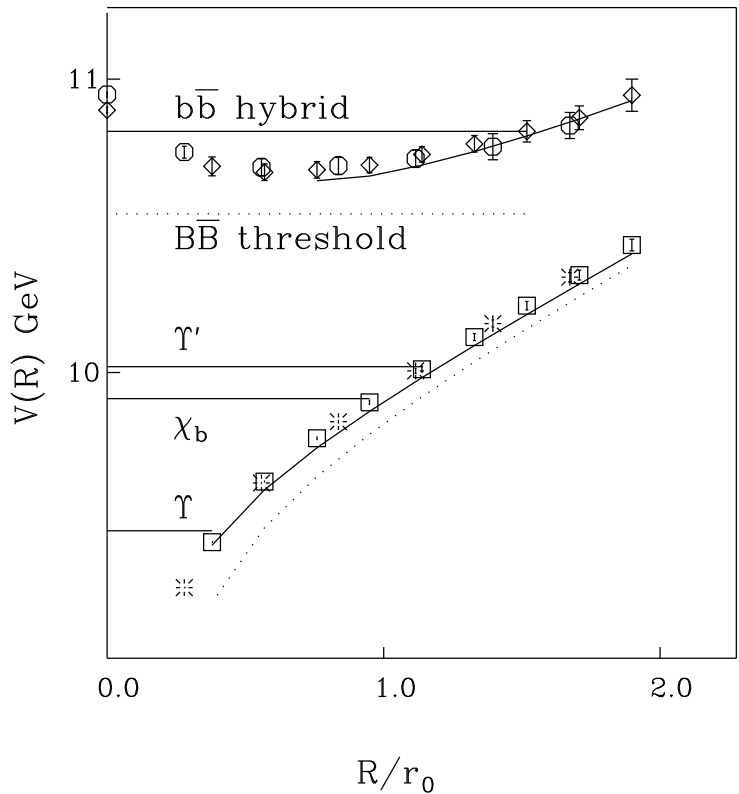}
 \caption{ Potentials $V(R)$ between static quarks  at separation $R$
(in units of $r_0 \approx 0.5$fm) for the  ground state ($\Box$ and {\tt
*}) and for the $\Pi_u$ ($E_u$)  symmetry which corresponds to the first
excited state of the gluonic flux (octagons and diamonds). Results  from
quenched calculations{\protect\cite{pm}} with SU(3) colour are shown by
symbols corresponding to different lattice spacings. The $R=0$ data
points{\protect\cite{glmsf}}  for the excited state are discussed in the
text. For the ground state  potential the continuous curve is an
interpolation of the lattice data while the dotted  curve with enhanced
Coulomb term fits the spectrum and yields the masses shown. For the
excited gluonic potential, the  continuous curve shows the string
excitation  expression of eq.~1. The hadronic bound states of $b$ quarks
in these  potentials are shown, including the lightest hybrid level in
the excited gluonic potential.
   }
\label{veu}
\end{figure}

On a lattice with the separation axis along a lattice axis, the relevant
 symmetry group is $D_{4h}$ and its representations are labelled such
that  $A,\ E $, and $B$   are related to the $J_z=0,\ 1$ and 2
excitations., etc. The lattice data for the lowest gluonic excited
states have been determined by the Liverpool group using the Wilson 
gauge action for SU(2) colour~\cite{liveu,pm89,heleu} and for SU(3)
colour\cite{pm}.  The lowest non-trivial  excitation was found to be
in the $E_u$ ($\Pi_u$) representation  (corresponding to flux  states
from an  operator which is the difference of U-shaped paths from quark
to antiquark of the form $\, \sqcap - \sqcup$). Although on a lattice,
the self energy of the static  sources is unphysical, the energy
differences between excited gluonic  potentials and the ground state are
physical and have a continuum limit. An illustration of this first
gluonic excitation is shown in fig.\ref{veu}.

Recently, a comprehensive study of this area has been undertaken using 
an asymmetric lattice~\cite{mornpear}. The advantage of using a
relatively  large spatial lattice spacing is that large $R$ can be
reached, while a smaller  temporal lattice spacing allows a more precise
determination of the energy values. The disadvantage of a coarse spatial
lattice spacing is  that lattice corrections are increased but this can
be offset by using improved  lattice discretisations with smaller
corrections than the Wilson lattice action which has order $a^2$. Recent
results extend out to $R$-values  of 4fm which is most impressive. By
using several different lattice spacings  and also by considering source
separations in other directions than the lattice axes,  it is possible
to extract continuum results for the excited gluonic potentials. 
Because of the lattice dependence of the self-energy, only energy 
differences can be obtained. 

\begin{figure}[ht!] 
\begin{center}
\epsfxsize=400pt\epsfbox{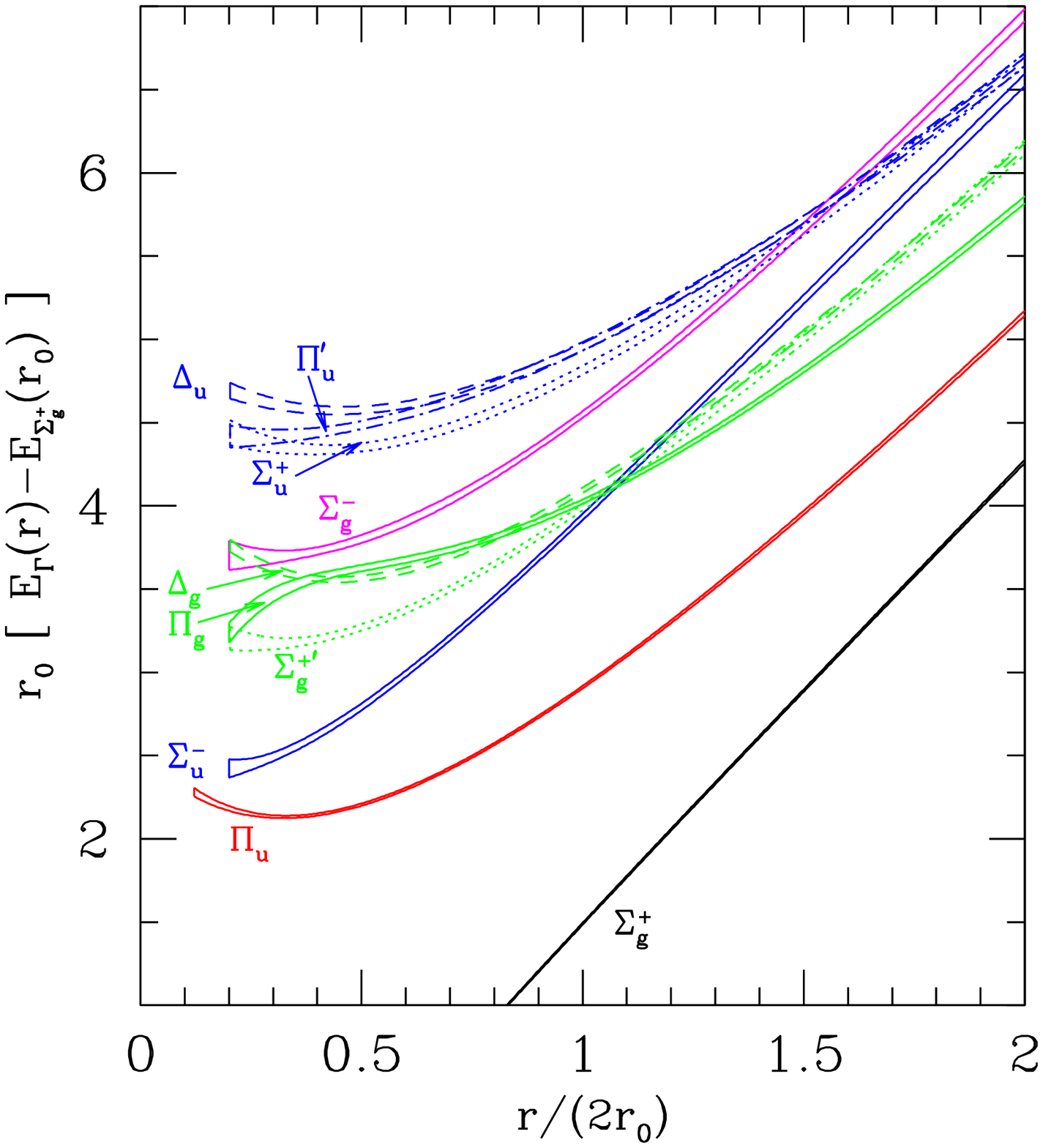}
\end{center}
 \caption{ Potentials $E(r)$ between static quarks  at separation $r$
(in units of $2r_0 \approx 1.0$fm) for different symmetry representations
 from Juge et al.~{\protect\cite{morn}} The two lines shown in each case
represent an  estimate of the error on the determination of the
continuum value.
   }
\label{mornfig}
\end{figure}

In order to set the scale in  lattice  calculations, it is now
conventional to use $r_0$. This is defined~\cite{sommer}  in terms of
the force between static sources  implicitly as $r_0^2 f(r_0)=1.65$
where $f(r)=dv(r)/dr$. From relating the  interquark potential to that
needed phenomenologically for heavy quark  bound states, we know that
$r_0 \approx 0.5$fm. Since this  potential can be very accurately
measured on a lattice, this is  an appropriate observable to use to set
the energy scale for comparisons of lattice results.
  Thus the lattice excited potential $\tilde{V}(R)$ expressed  as $R_0
(\tilde{V}(R)-V(R_0))$ where $R_0=r_0/a$ is  dimensionless and, in the 
limit of small lattice spacing, will be  equal to the continuum 
expression $r_0(\tilde{v}(r)-v(r_0))$ up to corrections of order $a^2$ 
for the Wilson lattice discretisation. This is thus a suitable way to 
present and compare results from different lattice studies. We present 
some of the recent comprehensive results~\cite{morn} in
fig.~\ref{mornfig}.

It is worth exploring what is  to be expected for these gluonic
excitations between static sources at separation $R$. 

\vspace{0.45cm}

{\em Stability}: The most obvious criterion is that if the excited 
gluonic energy lies below the energy of all states with the same symmetry
made from  a glueball together with a lower-lying potential, then the
excited gluonic potential will be  stable, at least in the quenched
approximation. Since the lightest glueballs are found, in quenched QCD, 
to have energies~\cite{ukqcd,gf11}  $m(0^{++})r_0=4.33(5)$ and
$m(2^{++})r_0=6.0(6)$, this criterion is satisfied in most cases,
especially  at larger $R$. 

\vspace{0.45cm}

{\em String models}: We consider now simple  phenomenological models for
the excited gluonic potentials:  one such model is the hadronic string. 
 A bosonic string fixed  at two points $R$ apart and with  no intrinsic
width will give a parameter-free prediction although the  tachyonic
problem implies that this prediction will be unphysical for  the lowest
energy excitation at small string length $R$.  The lattice excited
potentials can be compared with this bosonic  string model provided an
appropriate expression~\cite{pm89}  is used for excited level $n$:
 \begin{equation}
  V_n(R) = (\sigma^2 R^2 - {\pi \sigma \over 6} +2 \pi n \sigma)^{1 \over 2}.
 \end{equation}
 If one uses this  expression for the energy differences of excited
levels from the ground state, there are no free parameters since the 
string tension $\sigma$ is determined by the ground state potential with
 $n=0$. At large $R$, this expression has the simple consequence that 
gluonic excitations are at multiples of $\pi/R$ in energy higher  than
the ground state.  The relationship~\cite{pm89,pm,morn} between the
excitation symmetry and $n$ is that the $n=0,$ 1  levels are  given by
$\Sigma^+_g$, $\Pi_u$, while $n=2$ excites  $\Sigma^{+'}_g$,\ $\Pi_g$
and $\Delta_g$,  and $n=3$ excites $\Sigma^{\pm}_u$,\, $\Pi'_u$ and
$\Delta_u$.

In order to make the string  expression well behaved at small $R$,  the
string has to be regulated in some phenomenological way related to  its
intrinsic width - for instance  as in the Isgur-Paton flux tube
model~\cite{ip} or in the Hungarian bag model~\cite{hbag}.
 Nevertheless,   the  expression of eq.~1  was found to agree quite
well with lattice spectra obtained  for $R$ above 0.5fm  for both
SU(2)~\cite{pm89} and SU(3)~\cite{pm}. Indeed the curve drawn  in
fig.~\ref{veu} for the $\Pi_u$ excited potential is just that given by
the expression of eq.~1 for the difference from the ground state
($\Sigma_g^+$). 
 Comparison with the larger $R$  lattice data now
available~\cite{morn,allen} again gives good qualitative  agreement at
moderate $R$ values of 1 to 2fm. The ordering at the largest $R$
available (2 to 4fm)  agrees very well with that from the string
excitation approach but the energy  differences  agree less  well than
they do below 2fm.

This is surprising since the string  model should be most reliable at
the largest $R$. However, it should be kept in  mind that the systematic
errors in the lattice determinations~\cite{morn} are  largest at large
$R$ - coming both from the difficulty of separating the  many excited
states of a given symmetry which lie close together and  from possible
finite lattice spacing  effects since a coarse spatial lattice spacing
is used to reach large $R$. A further possible source of error is
from the  transverse size of the lattice which is assumed to be much
larger than the string length $R$ in the  string model whereas it is
comparable to  $R$ in some of the lattice analyses. So, it is possible
that the systematic  errors in the lattice results are somewhat
underestimated. This comparison between the hadronic string model  and
the large $R$ excited potentials is  discussed in  detail  by
Morningstar~\cite{morn}.

\vspace{0.45cm}

\begin{figure}[t]
\epsfysize=1.5in
\centerline{\epsfbox{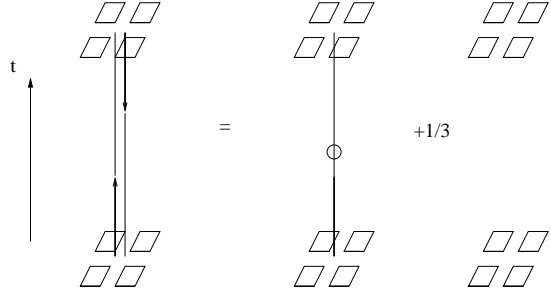}}
\caption{Relation of $R \to 0$ Wilson loop with adjoint source (gluelump) 
and glueball correlations}
\label{fig:wils}
\end{figure}

{\em The limit} $R \to 0$: At $R=0$, the symmetry considerations are 
different and this imposes constraints on the excited gluonic spectra 
that help to explain some of the more significant departures from the
string model. In the limit as $R \to 0$, the static source and
anti-source will be at the same  site and hence their colour can be
combined in a gauge invariant way - creating an adjoint colour source
and a  singlet (glueball correlator).  Pictorially this is shown in
figure~\ref{fig:wils}.  This former situation has been studied
previously~\cite{gl,glmsf}: the gluelump is an adjoint source in the
presence of a gluonic field - it is rotationally invariant and is 
described by $J^{PC}$.  Thus the Wilson loop correlation satisfies

 \be \lim_{R \to 0} W(R,t) = c e^{-M_{\rm gluelump }t}+\frac
{1}{3}c'e^{-M_{\rm glueball }t}\ee

\noindent In the large $t$ limit, the lighter of the two states will
dominate the  correlation function.  In most cases of present interest,
the gluelump state  is lighter than the glueball.  Thus, we can obtain a
relationship between the gluonically-excited states of the generalised
Wilson loop, in the limit $R \to 0$,  and those measured in the gluelump
spectrum. This relationship in the continuum is obtained by subducing
the rotation group  representations appropriate to the  gluelump  to the
$D_{\infty h}$ representations appropriate to the generalised Wilson 
loop when $R \ne 0$.  By subducing the irreducible representation  of
the gluelump with $J^{PC}$ we will find $D_{\infty h}$  representations
with $J_z= -J, \dots, J$; labels $g,\ u$ given by $CP$  and, for any
$J_z=0$ states an additional label given by $P (-1)^J$.  These
relationships are given in Table~\ref{tab:sub}.

Similar identities as $R \to 0$ also apply  to the lattice
discretisation. Then the   $O_{h}$ representations appropriate for the
gluelump can be subduced into the  $D_{4h}$ representations appropriate
 for the  generalised Wilson loop with $R \ne 0$.  Thus, as has been
emphasised previously~\cite{gl},   the ground state gluelump with 
$1^{+-}$ ($T_1^{+-}$)  implies that as $R \to 0$ there must be a
degeneracy of the  two-dimensional  $\Pi_u$ state ($E_u$) and a
$\Sigma_u^-$ state ($A_{1u}$). This  explains why the $\Sigma_u^-$ 
state which is high lying at large $R$ (since it is a $n=3$ string
excitation)  is seen in fig.~\ref{mornfig} to decrease in energy
rapidly at small $R$ since it has to be  degenerate at $R=0$ with the
$\Pi_u$ state which is  a $n=1$ string excitation and  so is lower
lying in energy. The lowest lying gluelump states~\cite{glmsf} have
been determined to  be $1^{+-},\ 1^{--}$ and $2^{--}$ with energy
differences in units of $r_0$ of 0.93(2) and 1.44(3) from  the $1^{+-}$
respectively. This implies that the $R=0$  limit of the excited gluonic
potentials should have degenerate states  as given in 
Table~\ref{tab:sub} with energy differences as  quoted above.  These
considerations  explain why the $\Sigma_g^-$, which is a $n=4$  string
level, also departs significantly (see fig.~\ref{mornfig}) from the
string-like expectation  since it has to be degenerate  at $R=0$ with
the $n=2$ string level $\Delta_g$.

\vspace{0.45cm}

{\em One gluon exchange}: Although the above group-theoretical
identities are a good guide to the behaviour of  the excited gluonic
potentials at small $R$,  the limit as $R \to 0$ of the excited gluonic
potential is not trivial  to extract from lattice data with $R=a,\ R=2a,
\dots$. A guide is  to consider the gluon exchange contributions
perturbatively. One way to  investigate this is to consider the self
energies of the contributions:  $2E_F$ at $R \ne 0$ and $E_A$ at $R=0$,
where $F$ and $A$ label fundamental  and adjoint colours.  Since, to
lowest order, $E_A= 9 E_F /4$ for SU(3) of colour, there will be a
mismatch and one might  expect the energy to increase as $R \to 0$ since
the adjoint self-energy is larger.  Another way to investigate this,  is
to imagine that as $R \approx 0$, there is a gluonic field in the
adjoint representation,  so that the heavy quark and anti-quark are also
in an adjoint and hence will have a Coulombic interaction energy given
by $-1/8$ of the Coulombic energy  between a quark and antiquark in the
fundamental representation (which is approximately given by $-0.25/R$ 
in lattice quenched studies). This again suggests that the excited
gluonic potentials should  rise as $R \to 0$, here as $0.03/R$. 

Lattice data for the $E_u$ representation for  small $R$ from SU(2)
colour studies~\cite{heleu} at $\beta=2.4$ with values of
$aV_{Eu}(R)=1.31$, 1.32 and 1.38 for $R= 3a,\ 2a$ and $a$ respectively
do  qualitatively support these estimates and are consistent with a
limit  as $R \to 0$ which agrees with the lattice gluelump
energy~\cite{adjbreak} of $aE_{\rm gluelump}= 1.50$. Also in
fig.~\ref{veu}, we show the $R=0$ point from the SU(3) gluelump
analysis~\cite{glmsf} which fits in well  with the above
considerations.

\begin{table}[h]

\caption{Connection between gluelump and two-body potential as $R \to 0$.}
  \label{tab:sub} \begin{center} 
 \begin{tabular}{cc} &\\ 
 \hline & \\ 
 Gluelump $J^{PC}$ &  Two-body potential states  \\  
  $1^{+-}$ &   $\Sigma_u^-$,$\Pi_u$\\
  $1^{--}$ &   $\Sigma_g^+$,$\Pi_g$\\
  $2^{--}$ &   $\Sigma_g^-$, $\Pi_g$, $\Delta_g$\\
 &\\ \hline

\end{tabular}
\end{center}
\end{table}

  It is possible to measure the distribution of the colour flux around 
an excited gluonic state. Results have been presented~\cite{helcol}  for
the $\Pi_u$ and for the first excited $\Sigma^+_g$ states. The  excited
states show a wider transverse distribution than for the ground state,
as would be expected.  There is some  evidence for a node in the 
transverse distribution of the first excited $\Sigma^+_g$ state.

\subsection{Hybrid Mesons with Heavy Quarks}

The  potential $V(R)$ between static sources obtained from  the lattice
(it is approximately of the form $V(R)=-e/R + \sigma R$) can be used to
determine the spectrum  of $b \bar{b}$ mesons by solving Schr\"odinger's
equation since the  motion is reasonably approximated as
non-relativistic. The  result from quenched lattices is similar to the
experimental $\Upsilon$ spectrum: see fig.\ref{veu}. The main 
difference is that the Coulombic part ($e$) is effectively too  small
(0.28 rather than 0.4). This produces\cite{pm} a ratio of  mass
differences $(1P-1S)/(2S-1S)$ of 0.71 to be compared  with the
experimental ratio of 0.78.  This difference is understandable  as a
consequence of the Coulombic force at short distances which would be 
increased by $33/(33-2N_f)$ in perturbation theory in full QCD compared 
to quenched QCD. Indeed lattice studies~\cite{df} including sea-quark 
effects in the vacuum do see evidence for an increase of the Coulombic 
component of the potential, although they have not yet reached sea-quark 
masses as small as those in nature.

 Consider now the hadrons corresponding to bound states of  the excited
gluonic potentials.  The lightest such  gluonic excitation ($\Pi_u$) 
corresponds to a component of angular momentum of one unit along the
quark-antiquark axis. Since the energy scale associated with the gluonic
 excitation is much larger than the energy scale associated with orbital
 or radial excitations, it is a reasonable approximation to use the
adiabatic approximation. Thus one solves for the spectrum of hybrid
mesons in the excited potential using the Schr\"odinger equation. For
those excitations  with a non-zero angular momentum $J_z$ about the
separation axis, an  extra centrifugal term has to be added. The spatial
wave function necessarily has non-zero angular momentum and 
corresponds~\cite{liveu} to $L^{PC}=1^{+-}$ and $1^{-+}$. Combining with
the  quark and antiquark spins then yields\cite{liveu} a set of 8
degenerate hybrid meson states with $J^{PC}=1^{--},\ 0^{-+},\ 1^{-+},\
2^{-+}$ and    $1^{++},\ 0^{+-},\ 1^{+-},\ 2^{+-}$  respectively. These
contain the  spin-exotic states with $J^{PC}=  1^{-+},\ 0^{+-}$ and
$2^{+-}$ which will be of special interest since  they do not occur as
$\bar{q}q$ states  and so any experimental evidence for a resonance with
these quantum numbers is a strong suggestion  for the existence of a
hybrid meson.

 Since the lattice calculation of the ground state and hybrid masses
allows  a direct prediction for their difference, the result for this
8-fold degenerate hybrid level is illustrated in fig.~\ref{veu}  and
corresponds\cite{pm} to masses of 10.81(25) GeV for $b\bar{b}$ and 
4.19(15) GeV for $c \bar{c}$. Here the errors take into account the
uncertainty in setting the ground state mass (i.e, $\Upsilon$ or $\psi$)
using the quenched potential as described above. Using the recent
comprehensive results\cite{morn}  confirms the  results above  and the
preliminary values quoted for the lightest hybrid meson is 10.8 GeV for
$b\bar{b}$ with no error estimate given.

 The quenched lattice results show that the lightest hybrid mesons lie 
above the open $B \bar{B}$ threshold and hence are likely to be
relatively wide  resonances. This could also be checked by  comparing
with quenched masses for the $B$ meson itself\cite{sommerb}, but at
present there are quite  large uncertainties on that mass determination.
The very flat potential implies a very extended wavefunction: this has
the implication that the wavefunction at the origin will be small,  so
hybrid vector states will be weakly produced from $ e^+ e^-$. An
explicit evaluation~\cite{morn} of the  hybrid wavefunction shows that
it is very extended:  being significant out to  a radius of 1fm.

 It would be useful to explore the splitting among the 8 degenerate
$J^{PC}$ values obtained. This could come from different excitation 
energies in the $L^{PC}=1^{+-}$ (magnetic) and $1^{-+}$
(pseudo-electric) gluonic excitations, from spin-orbit terms, as well as
from mixing between hybrid states and $Q\bar{Q}$ mesons with non-exotic
spin. One way to study this on a lattice is to use the  NRQCD
formulation which describes non-static heavy quarks which propagate 
non-relativistically. Preliminary results for hybrid excitations from
several  groups\cite{manke,collins,morn} give at present similar results
to those with the static approximation as described above,  but future
results may be more precise and able to measure splittings among
different states.

\subsection{Hybrid Mesons with Light Quarks}

 Unlike very heavy quarks, light quark propagation in the gluonic vacuum
sample is very computationally intensive --- involving inversion of huge
($10^7 \times 10^7$) sparse matrices. Current computer power is 
sufficient to study light quark physics thoroughly in the quenched 
approximation. The state of the art\cite{yoshie} is the Japanese CPPACS
Collaboration  who are able to study a range of large lattices (up to
about $64^4$) with a range of light quark masses. Qualitatively the 
meson and baryon spectrum of states made of  light and strange quarks is 
reproduced with discrepancies of order 10\% in the quenched approximation.

\begin{figure}[bt!] 
\vspace{11cm} 
\includegraphics{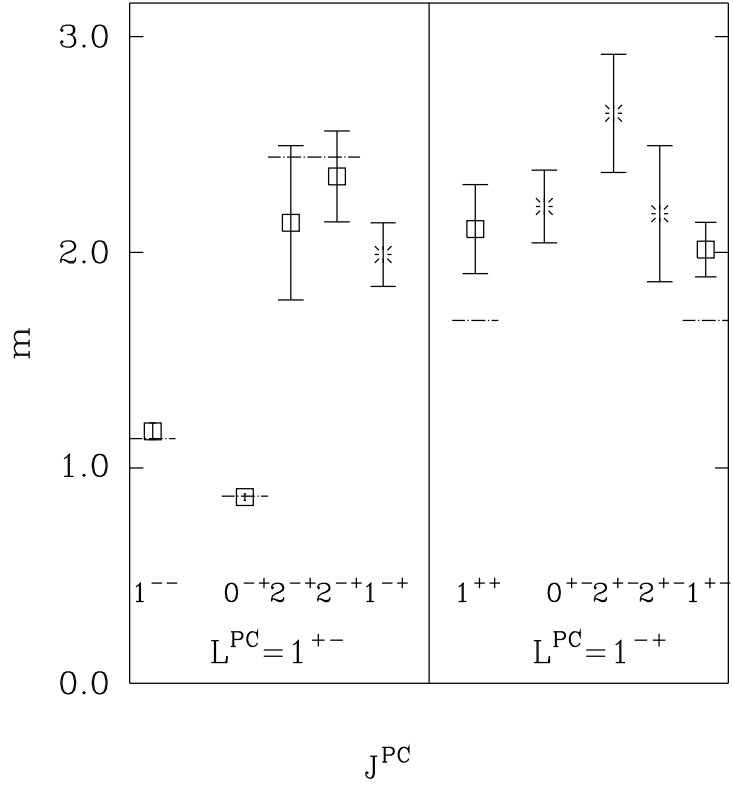}
 \caption{ The masses in GeV
of states of $J^{PC}$ built{\protect\cite{hybrid}} from hybrid
operators with strange quarks, spin-exotic ({\tt *}) and non-exotic
($\Box$). The dot-dashed lines are the mass values found for  $s\bar{s}$
operators.
   }
\label{hybf}
\end{figure}

 Here I will focus on hybrid mesons made from light quarks. In the
quenched  approximation, there will be no mixing involving spin-exotic
hybrid mesons  and so these are of special interest. The first study of
this area was by the  UKQCD Collaboration\cite{hybrid} who used
operators motivated by the  static ($Q\bar{Q}$) studies referred to
above. Using non-local operators, they studied  all 8 $J^{PC}$ values
coming from $L^{PC}=1^{+-}$ and $1^{-+}$ excitations. The  resulting
mass spectrum is shown in fig.~\ref{hybf} where the $J^{PC}=1^{-+}$
state  is seen to be the lightest spin-exotic state with a statistical
significance of 1 standard deviation. The statistical error on the mass
of this lightest spin-exotic meson  is 7\% but to take account of
systematic errors from the lattice determination, a  mass of 2000(200)
MeV is quoted for the $s \bar{s}$ meson. Although not  directly
measured, the corresponding light quark meson would be expected to be
around 120 MeV lighter. In view of the   small statistical error, it
seems unlikely that the $1^{-+}$ meson in the quenched approximation
could lie as light as 1.4 GeV where there are experimental  indications
for such a state\cite{hadron97}. There are also  recent experimental
claims~\cite{weygand} for a second $1^{-+}$ spin exotic meson at 1.6
GeV.  Beyond the quenched  approximation, there will be mixing between
such a hybrid meson and $q \bar{q} q \bar{q}$ states such as $\eta \pi$
and this may be significant in explaining the apparent discrepancy
between experiment  and lattice.

One feature clearly seen in fig.~\ref{hybf} is that non spin-exotic mesons
created  by hybrid meson operators have  masses  which are very similar
to those found when the states are created by $q \bar{q}$ operators.
This suggests that there is a significant  coupling between hybrid and $q
\bar{q}$ mesons even in the quenched approximation. This lattice result is 
difficult to quantify in terms of decay widths but does  imply that 
the $\pi(1800)$ is unlikely to be a pure hybrid, for example.

A second lattice group has also evaluated hybrid meson spectra from
light quarks. They obtain\cite{milc} masses with statistical and various
systematic errors for the $1^{-+}$ state of  1970(90)(300) MeV,
2170(80)(100)(100) MeV and 4390(80)(200) MeV for $n \bar{n}$,  $s
\bar{s}$ and $c \bar{c}$ quarks respectively. For the  $0^{+-}$
spin-exotic they have a noisier signal but evidence that it is heavier.
They also explore mixing matrix elements between spin-exotic hybrid 
states and 4 quark operators.

\section{The String Self-Energy}

\begin{figure}[tb]
\vspace{11cm} 
\includegraphics{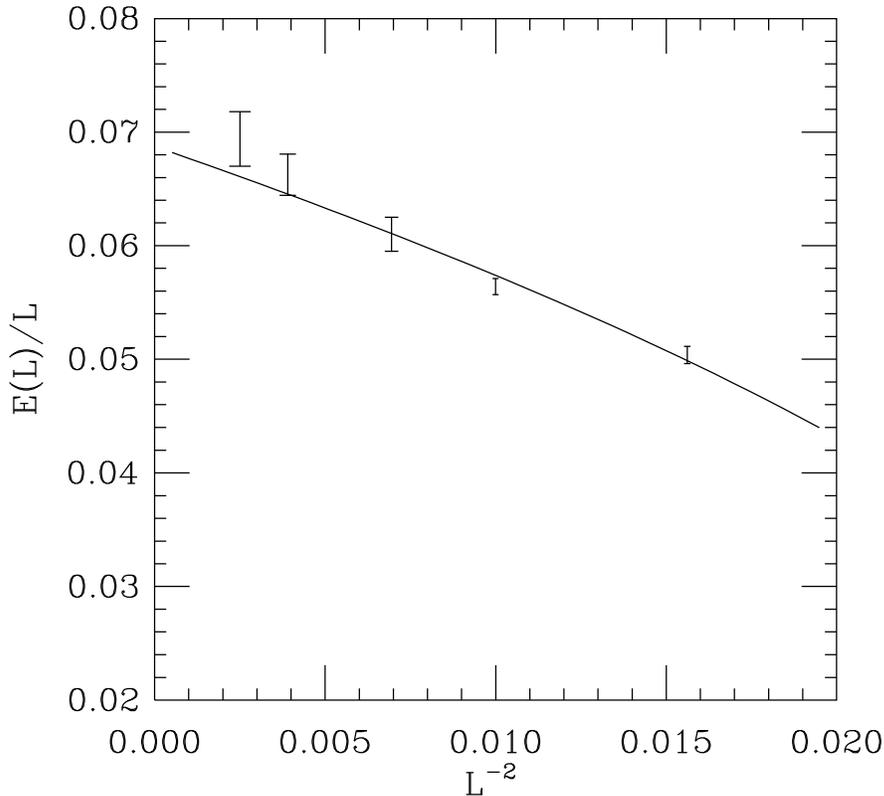}
 \caption{Relationship between energy $E$ and length $L$ around the
periodic  spatial boundary for a torelon in lattice
units{\protect\cite{cmpms}}. The  continuous line shows the bosonic
string result.
 }
 \label{torf}
\end{figure}

 The expression for the energy of  the ground state string mode ($n=0$)
will have  a contribution from  string fluctuation as well as the 
linear string tension component, namely\cite{lu} $V(R) \approx \sigma R
-\pi/(12R) $ as $R \to \infty$. In practice this  $\pi/(12R)$ string
fluctuation term for a bosonic string is very hard to disentangle from
the  Coulomb term $e/R$ in the potential. One way to get round this in
lattice studies is to consider a hadronic string that  encircles the
periodic spatial boundaries of length $L$. Then there are no  sources
and hence no Coulomb component. The appropriate string fluctuation term
in the  energy of this state, called the torelon, is then given by 
$E(L)=\sigma L - \pi/(3L)$ as $L \to \infty$. By  plotting $E(L)/L$
against $L$, see fig.~\ref{torf}, lattice studies\cite{cmpms}  have
confirmed the presence of this string fluctuation term with its 
coefficient determined to agree with the expected value and to  be
determined quite accurately (as $(\pi/3) \pm 0.03$). This is impressive
evidence that the hadronic string is a good model  of the energy of the
colour flux tube at large distances.

\section{Excited Colour Sources}

\subsection{Adjoint Sources}

\begin{table}[h] 
 \caption{Energy differences from the ground ($1^{+-}$) state}  of gluelump
states with $J^{PC}$. 
 \label{tab:continuum}

\begin{center} 
 \begin{tabular}{ccc} 
 &&\\
  \hline& & \\
  Excited State  &  $\Delta(M_0r_0)$  &  Energy (MeV)  \\
 && \\ \hline
 && \\
 $1^{--}$   & 0.933(18)    & 368(7)  \\
 $2^{--}$  & 1.438(25)   & 584(10)  \\
 $0^{++}$  & 2.771(72)   & 1092(28)  \\
 
 &&\\ \hline
 \end{tabular}
 \end{center}
\end{table}

 {\em One adjoint source}: The `hydrogen atom' of pure gauge QCD has a
static adjoint source  with a gluon field making it an overall colour
singlet. This  is the gluelump whose spectrum has already been mentioned
above. It has  been explored in SU(2)~\cite{gl} and 
SU(3)~\cite{glmsf} quenched lattice studies. The `magnetic gluon' with 
$J^{PC}=1^{+-}$ is found to be the ground state with the $1^{--}$ state 
as the first excited state. See table~\ref{tab:continuum}.  

 This spectrum would be accessible experimentally should a gluino  exist
which is sufficiently heavy for the static approximation to be valid and
 which lives long enough to allow spectroscopic energy levels to be
determined. 

 As well as exploring the spectrum, the distribution of colour fields
has been  measured~\cite{gl}. They are found to extend out to a radius
of about  0.5fm. This is consistent with the result (see below) that the
 adjoint string breaks into two gluelumps at a separation of around 1fm -
that is when  the two gluelumps are just touching. More detailed results
 give some evidence of the nature of the colour fields in the low-lying
states: disk-like for the $1^{+-}$ and toroidal for the $1^{--}$.

\begin{figure}[bt!] 
\vspace{11cm} 
\includegraphics{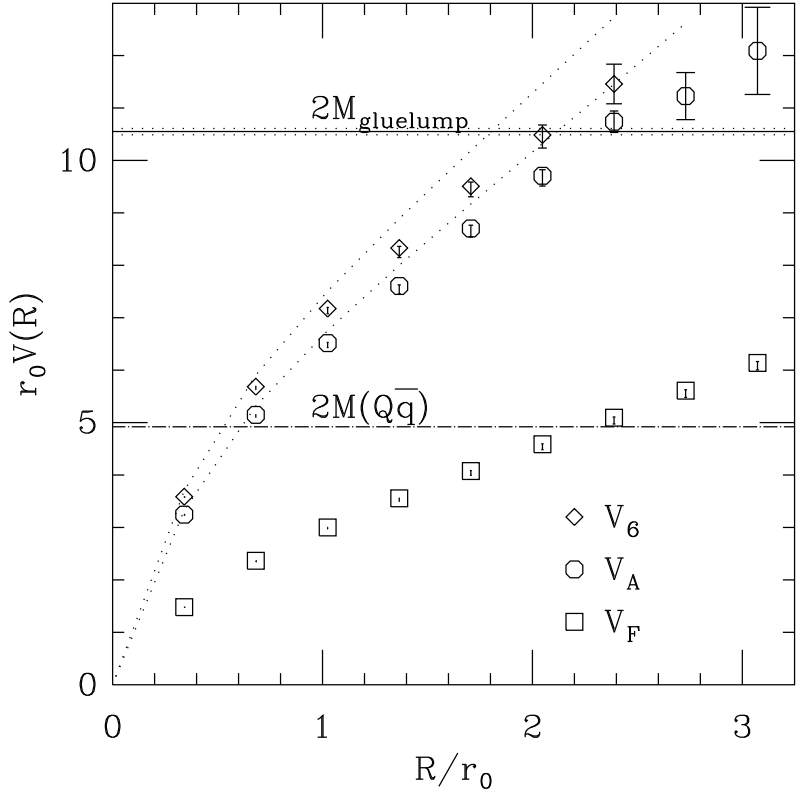}
 \caption{ The potentials in units of $r_0$ between static sources for
quenched SU(3) with fundamental, adjoint and sextet sources. From
quenched $\beta=5.7$  $12^3 24$ lattices, setting the scale using $r_0
\approx 0.5$fm.  
  Also shown\cite{glmsf,stok} are  $2M_{\rm gluelump}$ and
$2M(Q\bar{q})$ and the Casimir ratios ${9 \over 4}V_F(R)$  and ${5 \over
2}V_F(R)$.
   }
\label{v368}
\end{figure}

 {\em Two adjoint sources}: Another route to explore confinement is to
measure on a lattice the potential energy between  two static sources in
the adjoint representation of the colour group. Here three regions are 
expected as $R$ varies. At small $R$ one gluon exchange should give a
Coulombic region  with strength proportional to  $C_A/R$ where $C_A$ is
the Casimir appropriate to the colour  representation; at intermediate
$R$ an effective string tension may be  discernible, while  at large $R$
the adjoint potential $V_A(R)$  must become independent of $R$ because each
adjoint colour source can be screened by a gluonic field. Indeed at
large $R$, $V_A \to 2 m_{\rm gluelump}$ where the  gluelump is  the
ground state hadron with a gluon field around a static adjoint source
discussed above. Of interest to model builders is the adjoint potential
at  moderate $R$ values: the effective string tension region. Some
theories of confinement~\cite{who}  have suggested that the effective
adjoint string tension is  given by the Casimir  ratio (i.e.
$\sigma_A/\sigma_F = C_A/C_F$):  for instance  from assuming that
confinement in  four-dimensional QCD acts like two-dimensional QCD -
which has Casimir scaling exactly. 

  Precise data exist\cite{adjbreak} for $SU(2)$ of colour and they do show
a region of linear rise, although with a slope less steep than  that
given by the Casimir ratio (namely ${8 \over 3} V(R)$ where  $V(R)$ is 
the fundamental colour source potential discussed previously). 
 For SU(3) of colour, I have determined the adjoint potential 
to large $R$ and the result is shown in fig.\ref{v368}. Again the 
Casimir ratio, here ${9 \over 4} V(R)$, is an underestimate except 
at very small $R$ where the Coulomb contribution is expected to have 
this ratio. A rule that $V_A(R) \approx 2 V(R)$ seems to be a better guide 
to this intermediate $R$ region.

  The $R$ value where the two gluelump state  becomes degenerate in
energy with  the measured adjoint potential is found for both 2 and 3
colours to be  near 1.2fm. For SU(2) colour, a variational
basis~\cite{adjbreak} with both  colour-flux  and two-gluelump basis
states was used which allowed  the mixing  in this  string breaking 
region to be studied more readily.

Lattice studies have also been made\cite{trot} of the colour field 
distributions  for the adjoint potential. This allows a comparison of  
the distribution of the adjoint colour field with the fundamental case.
The longitudinal distribution at intermediate $r$-values shows a ratio
of field strengths falling below the Casimir  ratio so  in agreement
with the discussion above. The transverse distribution  is surprisingly
similar in the two cases - which may be  interpreted as giving evidence
that the two colour flux tubes involved in the adjoint case are
attracted to each other -  as in a type I superconductor.

\subsection{Sextet Sources}

 For SU(3) of colour, I have determined the sextet potential  to large
$R$ and the result is also shown in fig.~\ref{v368}. Again the  Casimir
ratio (here ${5 \over 2} V(R)$) is an underestimate except  at very
small $R$ where the Coulomb contribution is expected to have  this
ratio.  At large $R$, the sextet source can be screened by  gluons to
form an effective fundamental  source. Thus one expects the  sextet
potential to have a slope of the fundamental string tension $\sigma$ at
large $R$. There is  no sign of this, but the data are also consistent with
a change of  slope beyond 1.2fm.
 
\subsection{Fundamental string breaking}

 In passing, we point out that the fundamental potential itself, in full
QCD,  will show string breaking at  $R$-values where $V(R) >
2M(Q\bar{q})$ where by $Q\bar{q}$  we mean the ground state meson with a
light quark $q$ bound to a static heavy  quark $Q$. This comparison of 
the energies of the static potential and the two meson system can be 
explored in the quenched approximation - although there will be no
actual mixing between these systems in that case. The situation is 
illustrated in fig.~\ref{v368} and again breaking  would be  expected
to occur at  an $R$-value  near 1.2fm where there is a degeneracy. 
Attempts to look for this mixing explicitly using $V(R)$ determined from full
QCD  simulations~\cite{df,ukqcddf} have not yet achieved sufficient
precision to address  this topic.

\section{Multiple Sources}

\subsection{Baryonic Applications}

\begin{figure}[tb]
\vspace{5cm}  
\includegraphics{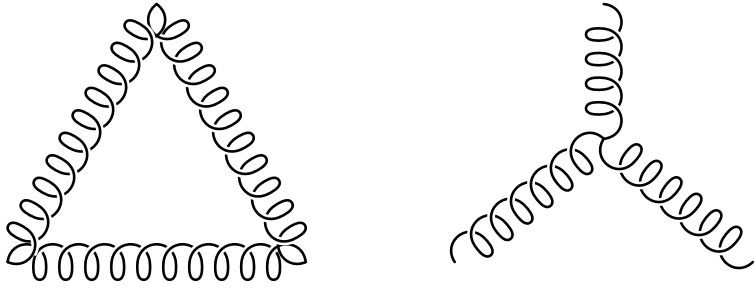} 
 \caption{Diagrams showing two possible assignments of colour flux 
between three static quarks in a baryonic configuration: as a sum of
two-body  terms or as a minimal length string.
 }
\label{barf}
\end{figure}

 Early attempts to study the potential with three static quarks were 
rather exploratory~\cite{baryonold}.  Indeed at moderate $R$ values, 
the baryonic potential was found to be given quite well by an average 
of the three two-body potentials.  In contrast, in a flux tube approach,
 one would expect an important contribution from a  three-body term with
a   flux tube of minimal length (ie with a star-like configuration): see
 fig.~\ref{barf}.  This would  give a  different  behaviour of  the
energy levels as a function of the position  of the three sources than
the case of a two-body sum.  The above studies had rather small  interquark
separations of 0.5fm or less which may be an explanation for  the
apparent dominance of the two-body terms.
 Some recent preliminary results~\cite{baryon}  give some support to the
three-body  flux-tube picture for the special case of 3 sources at the 
corners of an equilateral triangle. 

\subsection{ $Q \bar{Q} Q \bar{Q}$ }

\begin{figure}[ht]
\vspace{5cm}  
\includegraphics{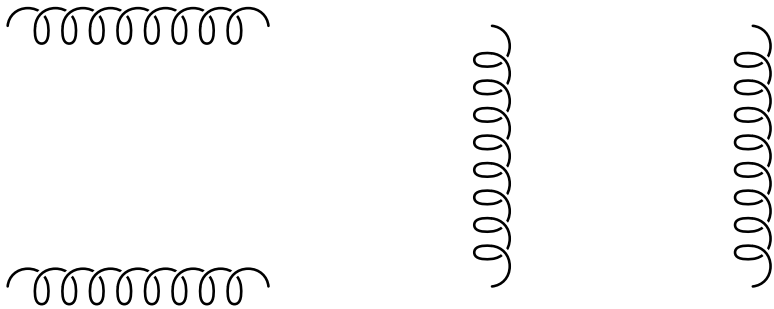} 
\caption{Diagrams showing two possible assignments of colour flux 
between  static $Q\bar{Q}Q\bar{Q}$.
}
\label{q4f}
\end{figure}

 To study the residual strong force between hadrons on a lattice is 
difficult since it is so much weaker than the colour force between
quarks. For  this reason, preliminary studies have either used static
quarks and/or SU(2) of colour to simplify the computations.  There has
been extensive work~\cite{qqqq} using static quarks in SU(2) colour
lattice  studies - this has a bearing on the meson-meson force. Of
special interest is whether the measured 4-body potential can be
expressed  as a sum of 2-body components. It turns out that a description
in  terms of the mixing of different 2-body assignments as shown in
fig.~\ref{q4f} is possible but that  the mixing coefficient is
geometry-dependent. In other words there is  a four-body force.

 As well as the energy levels for a wide range of geometries, the spatial 
distribution of the colour flux has been determined for the case 
of 4 sources at the corners of a square~\cite{q4col}.

\subsection{The BB interaction}

\begin{figure}[h]
\vspace{5cm} 
\includegraphics{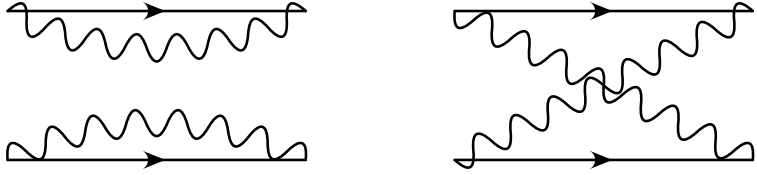} 
\caption{Diagrams showing the interaction between two B mesons: the light 
quarks are shown as wiggly lines.
}
\label{bbdf}
\end{figure}

\begin{figure}[tb]
\vspace{11cm} 
\includegraphics{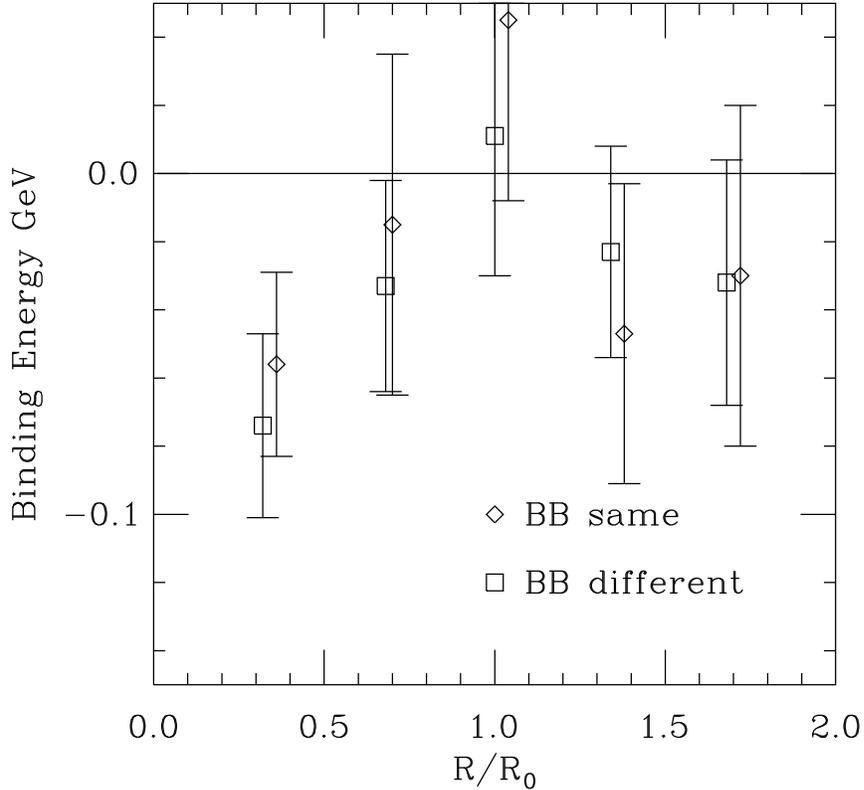}
 \caption{Preliminary results~\cite{BB} for the binding energy  between
two $B$ mesons with either same flavour  of light quarks, or different
flavours at separation $R$ in units of $R_0 \approx 0.5$fm. The light
quark mass used corresponds to strange quarks. At $R=0$, the binding
energy  is given from lattice results for  baryons with static
quarks~{\protect\cite{stok}} as around -0.3 GeV  for different flavours
and -0.2 GeV for same flavour- see the text.}
 \label{bbf}
\end{figure}

The $Q\bar{q}Q\bar{q}$ system with the heavy quarks $Q$ treated  as
static at separation $R$ is a step towards studying the  strong force
between realistic hadrons.  This case can be considered as the
interaction between two B mesons in the heavy quark limit -  where the
$b$ quarks are taken as  static. With the B mesons  separated by
distance $R$, the interaction comes from  the mutual influence of the
light quarks. The relevant diagrams to  study are shown in
fig.~\ref{bbdf}. A determination of this potential  energy will allow a
non-perturbative determination of the spectrum of BB  bound states -
which are exotic states of considerable interest.

Exploratory studies had been made for the cross diagram only  for SU(3)
colour~\cite{BBold} and for both diagrams  using SU(2)
colour\cite{BBcanada}. Recently a preliminary study has been made  using
 SU(3) colour for  quenched lattices~\cite{BB} with light quarks of 
mass near the strange quark mass. Only if the two B mesons
 each have the same flavour of light quark will the   additional cross
diagram contribute.  Preliminary results  for  same light flavour and
different light flavour are shown  in fig.~\ref{bbf}. Binding is seen in
each  case, but it is  stronger  in the unlike flavour case. This
suggests  that a $B-B_s$ exotic dimeson might exist. It will also be
possible  to study the spin-dependence of these potential energy
differences on a lattice.

 Note that again, at $R=0$, the system is related to one previously
studied.  Thus  the colour of the two static quarks can be combined to 
a triplet or sextet and the lightest state will  have them in a colour
triplet - creating a system like  the $\Lambda_b$ baryon but with a
static  $b$ quark. These baryonic states with one static quark have
been studied  on the lattice previously and their mass   gives the BB 
energy on the lattice at $R=0$. Lattice estimates~\cite{stok,BB} give
around 0.3 GeV  for this binding at  $R=0$. For light quarks of the
same flavour, there is no $\Lambda_b$  state - the lightest baryon will
be the $\Sigma_b$ which is somewhat heavier so  the binding in this
case will be around 0.2 GeV.

In the continuum, in the heavy quark limit, one would expect that the
binding of the BB system at $R=0$  for unlike light quark flavours is
given by  $2(M_B-m_b)-(M_{\Lambda_b}-m_b)$ where $m_b$ is the $b$-quark
mass.

The interaction energy between a  $B$  and $\bar{B}$ can be studied
similarly. However, this system at separation $R$ has a coupling to a
state consisting of the flux tube joining the static sources  (ie the
state which occurs in the static potential $V(R)$) plus a light
quark-antiquark meson (or vacuum when the light quark and antiquark have
 identical flavour).  This reflects the physical process that  $B
\bar{B} \to \Upsilon + q \bar{q}$ and since the $\Upsilon$ is known to
be  tightly bound, this latter channel may have a lower energy at small
to moderate $R$ values. Conversely, at large $R$, this transition  is
essentially that of fundamental string breaking: where it is 
energetically favourable for the string of length $R$ to break leaving 
a $B$ and $\bar{B}$. A thorough study of this area thus needs  to use  a
basis with both string (flux) states and meson-antimeson states.

\section{Conclusions}

 The lattice has a lot to offer to phenomenological model builders and those 
wishing to unravel the mysteries of QCD. 

 String-like features are seen in the excited gluonic potentials for  $R
> 1$fm. The departure from this behaviour at moderate and small $R$,
especially for the $\Sigma^-$ states, can be understood as a consequence
of the  symmetry requirements at $R=0$. There is even  lattice evidence
from closed string studies for the self energy appropriate to  a bosonic
string. 

 As well as predictions for hybrid mesons with heavy quarks, recent lattice 
results give predictions for the light quark hybrid sector. There  seems 
to be a discrepancy between the lattice expectation of around 1.8 GeV 
and the experimental candidate at 1.4 GeV.

 The response of the gluonic vacuum to static sources in adjoint 
and sextet representations was discussed. An effective adjoint 
string tension approximately double that of the fundamental was found. The 
string breaking for both adjoint and fundamental sources appears to occur 
for separation of around 1.2 fm. 

 It is possible to study the residual strong force between colour
singlet hadrons on a  lattice.   Results were presented for this 
interaction - both among 4 static  sources (ie like the $\Upsilon$ -
$\Upsilon$ system) and among  two heavy-light mesons(ie like the BB
system).

\end{document}